# IDENTIFICATION OF EFFICIENT PEERS IN P2P COMPUTING SYSTEM FOR REAL TIME APPLICATIONS

Jigyasu Dubey[1] and Vrinda Tokekar[2]

[1]Department of Information Technology, Shri Vaishnav Institute of Technology & Science, Indore, India
jigyasudube@yahoo.co.in
[2]Institute of Engineering & Technology, Devi Ahilya Vishwavidyalaya, Indore, India
vrindatokekar@yahoo.com

## ABSTRACT

*Currently the Peer-to-Peer computing paradigm rises as an economic solution for the large scale computation problems. However due to the dynamic nature of peers it is very difficult to use this type of systems for the computations of real time applications. Strict deadline of scientific and real time applications require predictable performance in such applications. We propose an algorithm to identify the group of reliable peers, from the available peers on the Internet, for the processing of real time application's tasks. The algorithm is based on joint evaluation of peer properties like peer availability, credibility, computation time and the turnaround time of the peer with respect to the task distributor peer. Here we also define a method to calculate turnaround time (distance) on task distributor peers at application level.*

## 1. INTRODUCTION

The Peer-to-Peer (P2P) distributed computing systems have attracted more and more research efforts recently. Currently several P2P networks have been developed for various types of applications such as distributed video encoding [1], file sharing [2][3], and media streaming [4]. When a P2P system is used for distributed computing great processing power can be achieved. Applications such as distributed.net[5], and SETI[6][7] uses the idle CPU cycles of thousands of computers connected to the Internet in order to break encryption codes and find signs of intelligent life in outer space. These systems are not decentralized systems, because they have just one single host distributing workload, whereas all other peers operate as workers. However, these applications have shown that the combined computational power of even a few hundred peers is often the only way to efficiently and economically solve very expensive computation task. To ensure high throughput and good utilization of resources it is required to have proper resource management. Due to the ad-hoc and dynamic nature of the P2P network paradigm, varying resource availability and unpredictable latencies are present which causes number of challenges in managing the computing resources and scheduling task execution across the systems [8]. Moreover the tasks in real time applications have deadline to be met which requires the predictable performance of the computing system [9]. The P2P computing systems have lack of scheduling schemes which clearly analyze the resource requirement and predict total execution time. Also they lack of mechanism for systematically recruiting the resources (peers) available in the system. To recruit the reliable peers some earlier systems have focused on identifying good peers [10], [11] based on the capability and reliability of peers. However they are not considering the communication overhead presents in P2P network which is major cause behind the time delay. It is very difficult to obtain accurate distance metrics for a dynamic, decentralized P2P networks. The number of peers involved in such systems is very huge; hence usual techniques such as *ping* and *traceroute* are not useful. The factor turnaround time reflects jointly processing capability of peer as well as communication overheads involved. In this paper we define a task unit and measure the distance between task distributor peer and task processors peers at the application level in the form of turnaround time of a task unit. We proposed a method to identify group of efficient peers in P2P computing System, to process real time applications tasks. The method identifies the reliable peers from existing peers, for processing of real time application tasks, on the basis of peer properties like peer credibility, peer computation time, and turnaround time. The peer credibility is the probability that the result produced by a peer is correct. Peer computation time represent the time when a peer actually executes the system's computations in the presence of peer autonomy failures.

## 2. RELATED WORK

P2P systems are different in both technological aspects and design/implementation issues. Recently, considerable research effort is being done on several important issues related P2P systems. In P2P computing networks, to minimize total processing time, Jingnan Yao and Laxmi Bhuyan in [12] designed a distributed packet processing algorithm known as resource sharing distributed packet processing algorithm known as resource sharing distributed packet processing algorithm (RSDLP). In RSDLP algorithm the workload is distributed among peers by organizing them into an efficient resource tree. Jingnan Yao, Jiam Zhou, et. al. in [9] proposed an load sharing mechanism for computing real time jobs in P2P networks. The proposed mechanism identifies most efficient resource pool with an optimized load scheduling. Author focus to explore maximum network utilization by building a resource tree of maximum efficiency (MET). Selection of peers is based on a combined evaluation of the available computation power and communication bandwidth. However, for peer selection, peer credibility is also required to be considered. Javier Celaya and Unai Arronategui in [13] proposed a new scalable scheduler for workflows with deadlines in a P2P desktop grid. The scheduler is built upon a tree-based network overlay with a distributed management of availability time intervals of resources. In proposed scheduler, authors only consider the availability time intervals of resources, however for the deadline driven tasks other parameters like resource credibility, actual participation time, and communication bandwidth also need to be considered. Virginia Lo, et al. in [14] proposed a system named cluster computing on the fly (CCOF) which harvest the CPU cycles from ordinary users (Desktop PCs). They also proposed a wave scheduler which exploits the large blocks of ideal time at night, to provide higher quality of service for deadline-driven jobs, using a geographic based overlay to organize hosts by time zone. In this wave scheduler they explore the possibilities to capture the CPU cycles from number of machines that lie completely idle at night. It provides a higher guarantee of ongoing available cycles hence it is useful for deadline driven tasks. The system provides the higher computation performance but due to using the peers from same night time zone which belongs the same geographic location the reliability of the system decreases. Dimitris Kamenopoulos, Iosif A Osman, et.al. in [15] presents a simple distance measurement method in volunteer computing networks. The method is based on passive monitoring of application level traffic. For the purpose of load balancing authors suggest that the distances between peers in a volunteer computing network need to be considered. In [16] authors proposed an algorithm to identify the reliable peer groups in P2P computing systems by using the peer properties like peer availability, credibility and computation time. However for the real time applications peer distance also be considered.

To identify the reliable peers for in P2P computing systems some earlier systems have focused on good peers [10], [11] based on capability and reliability of peers where as some have focused on distance value between peers. The real time and scientific applications also demand accurate results. For such applications, in P2P computing systems it is required that peer selection is based on peer credibility along with the above discussed peer properties.

## 3. P2P COMPUTING SYSTEMS - TOPOLOGY & CHARACTERISTICS

The P2P computing systems utilize the processing power of idle Desktop PCs presented at the edge of the Internet as shown in figure 1. The P2P computing systems support only those computations which can be divided into small tasks and which are embarrassingly parallel in nature. The P2P computing systems consist of two major components- Task Distributor and Task Processor [17]. A peer in a P2P computing system at a time can be a task distributor or a task processor, but not both. A user on a peer provides the job in form of computation code with the data which is to be processed. This peer becomes as a task distributor in the system and other peers as task processors. The task which is provided by the user for processing must be able to split in subtasks. The task distributor peer is responsible for splitting the task in to number of small tasks and distributes these subtasks to the task processor peers in the system for processing. The task distributor is also responsible for integrating the results which it receives from the different task processors after the completion of task. The task processor peers receive the computation code and data from the task distributor and return the results back to the task distributor after processing the task.

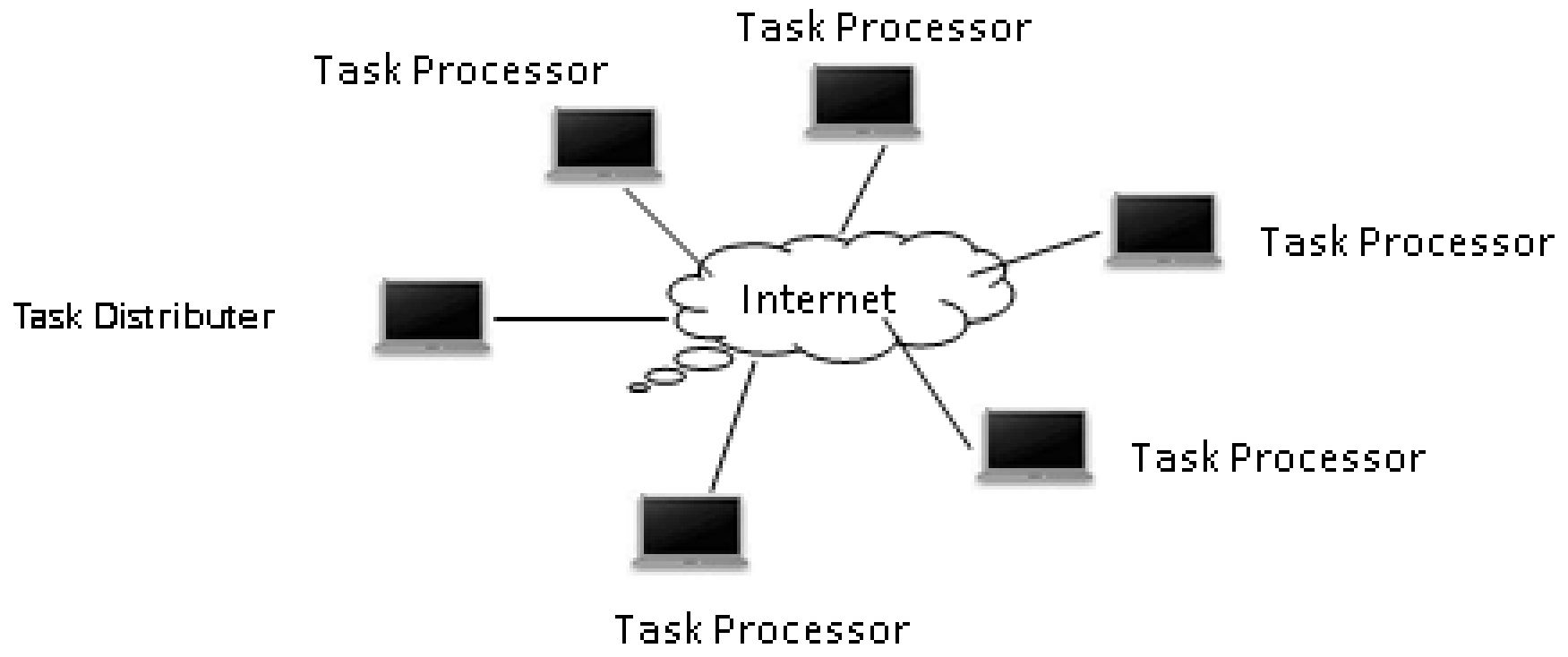

Figure 1. P2P Computing System Model

The P2P computing systems act like virtual super computer with varying processing power by utilizing the ideal CPU cycles of the desktop PCs connected to the Internet [18]. The number of peers in a P2P computing system may range from less than 10 to hundreds of thousands. The pure P2P computing system allows any peer to act as a task distributor or a task processor. For the successful and efficient working of system, it is required that in a P2P computing system very few peers act as task distributor, whereas the majority of the peers act as task processors. It is expected and required that most of the peers in P2P computing system will take the responsibility for processing the task, otherwise the computation take very long time. It is an important characteristic of P2P computing system because it implies that the topology of P2P computing system is primarily a forest like structure, where one-to-many type of communication occurs.

The figure1 describes an abstract model of P2P computing system which consists of heterogeneous peers interconnected via heterogeneous communication links. Each peer in the diagram is capable of task processing. In P2P computing system peers can have different processing power and they are interconnected with each other in an arbitrary fashion via links with different bandwidths. Each peer has its own local task processing in the background. The above discussed conditions lead towards delay in task completion time, which is intolerable for real time applications. It also indispensable that result produced by each peer is correct. Due to the dynamic nature of peers and available network latencies it is very difficult to use this type of systems for computation of real time applications. The scientific and real time applications require predictable performance because tasks in such applications have dead line to be met.

In this paper we proposed an algorithm to identify group of efficient and reliable peers from the existing peers in P2P computing systems, for processing of real time application tasks. The proposed algorithm is based on joint evaluation of peer properties: peer credibility, computation time, and distance of the peer.

### 3.1. Peer Credibility

The peer credibility $C_P$ in [13] given as the probability that the result produced by a peer is correct.

$$C_P = C_R / (E_R + C_R + I_R) \qquad (1)$$

Here, $E_R$ represents the number of erroneous results, $C_R$ represents the number of correct results, and $I_R$ represents the number of incomplete results. The term $E_R + C_R + I_R$ represents the number of total tasks that a peer computes.

### 3.2. Peer Computation Time

The peer computation time ($PC_T$) in [13] given as the expected Computation time when a peer processes the system's computations during IT.

$$PC_T = IT \;\; X \;\; A_P \qquad (2)$$

Where $IT$ is peers ideal time and $A_P$ is peer availability. The Peer computation time represent the time when a peer actually executes the system's computations in the presence of peer autonomy failures.

### 3.3. Distance of Peers

The peers in P2P computing system works in a completely passive manner hence need no knowledge of distance information. For the purpose of load balancing, measuring any kind of distance between peers is unnecessary, because a peer becomes a task distributor only when actual workload occurs. The only parameter that needs to be considered is the distance between task distributor peer and corresponding task processor peers.

The P2P computing system is designed for those applications that have large life time. A real time task $T_i$ is characterized by its arrival time $\omega_i$, execution deadline $\omega_d$, and size $ST_i$. A job is considered successful if it can be executed before its deadline. In P2P computing system for the same application, individual tasks are usually approximately equal in size and need equal computational requirements. When a peer decides to become a task distributor and divide its workload in to number of smaller tasks say $t_1, t_2, t_3, t_4, \ldots, t_{n-1}, t_n$ , the task size has lower limit, beyond which further division of a application makes no sense. In P2P computing system we call this smallest possible task as "task unit". In P2P computing system a task unit has following properties-

- Any task assigned from a task distributor to a task processor consists of a positive integer number of task units. For a same application total number of task unit is represented by $n$.

- We represent the size of task unit by *ST*. In P2P computing system all the task units have equal size for a same application, i.e. $ST_1 = ST_2 = ST_3 = ST_4 = \ldots\ldots\ldots = ST_{n-1} = ST_n$.

- The number of task units ($n$) for a given application is much larger than the number of task processors ($T_p$) involved, i.e. $n >$ Number of $T_p$.

From the above discussion we found that the size of task unit (*ST*) in P2P computing system is constant and a static property. The distance between the task distributor and task processor is dynamic and to minimize processing time we take the advantage of this parameter. The distance $D\ (T_D,\ T_P)$, where $T_D$ is task distributor peer and $T_P$ is a task processor peer, can be defined as the time interval between the instance where $T_D$ dispatch a task unit to $T_P$ and the time the result of this task unit is returned to $T_D$. The distance $D\ (T_D,\ T_P)$ is the turnaround time for a task unit. The value of turnaround time $D\ (T_D,\ T_P)$ for a task unit highly depends on parameters like network bandwidth, task processor load and task processor peer's processing capability. Suppose task distributor peer ($T_D$) dispatches a task unit $t_{id}$ to a task processor ($T_P$) at time $T_S$ and receives result of $t_{id}$ at time $T_C$, than turnaround time $D\ (T_D,\ T_P)$ is given as:

$$D\ (T_D,\ T_P) = T_C - T_S \qquad (1)$$

For simplicity we assume that $T_D$ dispatches single task unit at a time to the $T_P$. But it may be the case that $T_D$ dispatches two or more task unit at a time to $T_P$. In this case, suppose $T_D$ dispatch $n$ task units to $T_P$ and received the cumulative result in time $D\ (T_D,\ T_P)$, the measured turnaround time for a task unit is $D\ (T_D,\ T_P)\ /\ n$. In P2P computing system a task distributor peer requires to keep a set of touples (Task processor ID, turnaround time) for the load balancing related decisions. There is no need to do anything on task processors, and also task distributor nodes does not need to exchange distance information between them.

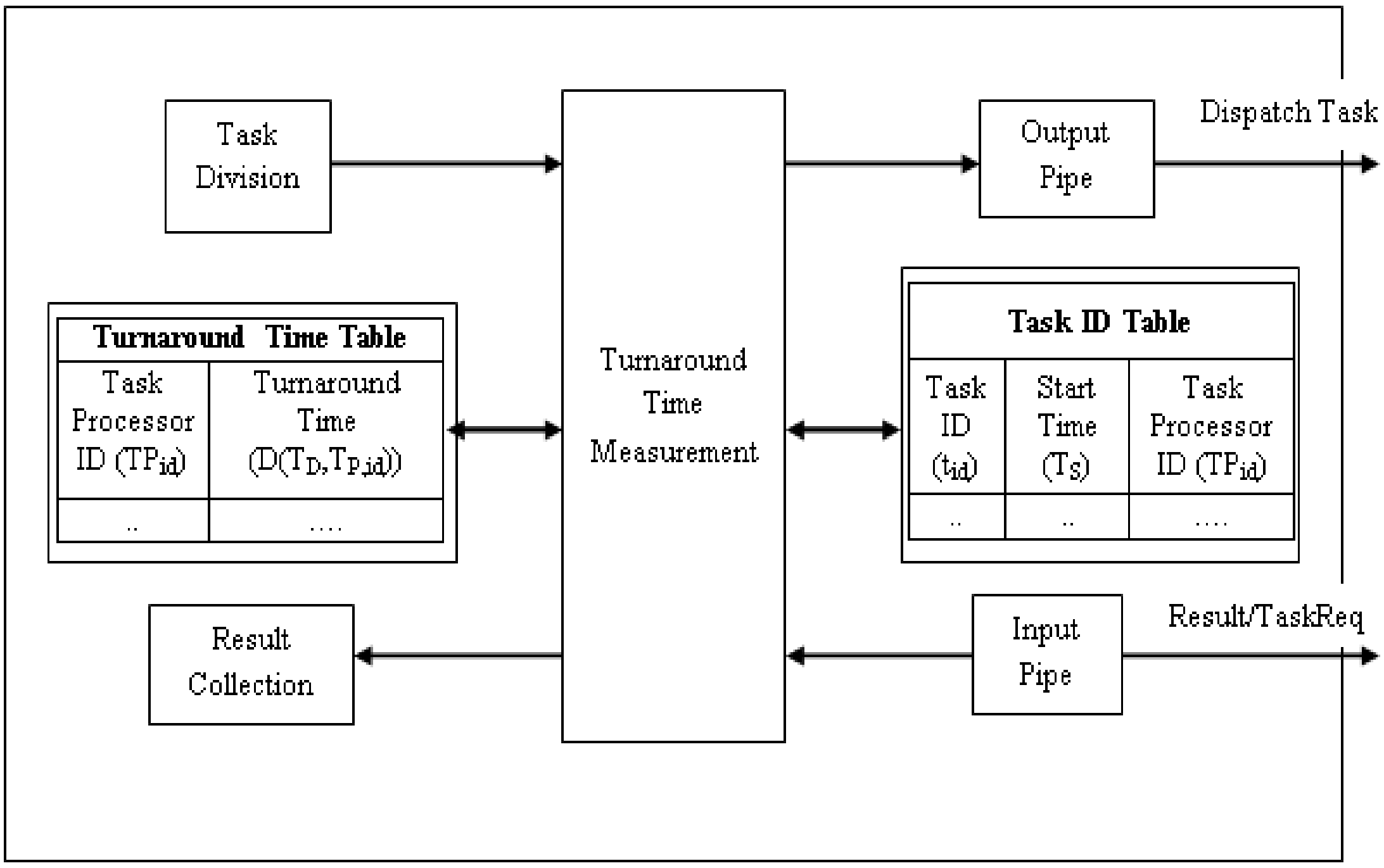

Figure 2. Task Distributor Model

On a conceptual level, each task distributor peer needs to maintain only a single lookup table named Turnaround-Time table as given in figure 2. The Turnaround-Time table contains task processor ids as key and the value of estimated turnaround time. Whenever a task processor peer returns a result, the corresponding distance value in the lookup table updated accordingly. To measure the task unit turnaround time, the task distributor peer must require to record the time when respective task is dispatched. In order to address this problem the task distributor peers maintain a second lookup table named TaskID table (shown in figure 2) containing the task ids, corresponding task start time and the task processor id to which task is dispatched. Whenever a result is returned, its task id is matched against the respective entry in the lookup table and the turnaround time is calculated by subtracting the time values stored there from the current time. The following algorithm calculates the turnaround time of peers on the task distributor peer.

**Algorithm for Distance Measurement on Task Distributor Peer**

// Turnaround time between $T_D$ & $T_P$ = D ($T_D$, $T_P$)

// Task Processor ID = $TP_{id}$

// Task ID = $t_{id}$

// Start time of $t_{id}$ = $T_S$

// End time of $t_{id}$ = $T_C$

// Number of subtasks represented by $t_{id}$ = n

1. Start

2. Launch the peer into JXTA network;

3. if (task is available for processing)

4. Divide the task into multiple subtasks;

5. Create input pipe advertisement;

6. Build input pipe by using the pipe advertisement;

7. Broadcast a pipe advertisement to announce the task availability;

8. while (subtask is available for processing)

9. Wait for response;

10. if (response == result)

11. Extract $T_S$ from TaskID table corresponding to $t_{id}$;

12. D ($T_D$, $T_P$) = $T_C$ – $T_S$;

13. if ($t_{id}$ represents more than one task)

14. D ($T_D$, $T_P$) = D ($T_D$, $T_P$) / n;

15. if ($TP_{id}$ is not available in turnaround-time table)

16. Insert $TP_{id}$ and corresponding $D(T_D, T_P)$ in turnaround-time table;

17. else

18. Update the value of $D(T_D, T_P)$ in turnaround-time table;

19. end else

20. Store the result;

21. end if

22. else

23. Create output pipe;

24. In TaskID table enter values of $t_{id}$, $T_S$ and $TP_{id;}$

25. Send the subtask and data to the peer which responds via output pipe;

26. end else

27. end while

28. if (no more results awaited)

29. Integrate all the results of subtasks;

30. end if

31. Go to step 3;

32. end if

33. end

## 4. IDENTIFICATION OF GROUP OF RELIABLE PEERS

A P2P computing system is based on desktop PCs which are connected to the Internet. In such systems it is assumed that each peer in addition of processing its local workload, has spare computing cycles to share with other peers. The real time application task that required computational resources, can initiate from any peer at any time with varying resource requirements and time deadlines. The spare computing power available on each peer varies and its availability time differs. The peers in the system can freely join and leave the system, in between the computation, without any constraints. Such situations are known as peer autonomy failure [13]. It leads to the delay and suspension of computations which are not acceptable for real time application tasks. To avoid such conditions it is required to identify group of most effective peers among the vast number of peers on the network, which process the real time application tasks. To identify such group, choice of peers not only depends on the absolute processing power that is available on a peer but also on the communication cost used to distribute the task and reception of task results. When a high processing peer is connected via an extremely slow link or when a fast link connects to a peer that is hardly available, it is required not to use such peers in processing of real time application tasks. Thus, it may not be very helpful for real time application task, to blindly include those peers that have either maximum processing power or minimum communication bandwidth, in P2P computing system. Our aim is to identify the most effective peers among the lots of peers available on the Internet, for the successful execution of real time application task. In this section we proposed an algorithm to identify the group of most efficient peers from the existing peers in P2P computing system.

From the above discussion, it is clear that performance of a P2P computing system is strongly dependent on task execution time (Processing power of peer) as well as communication cost. The successful execution of task on a peer is strongly relying on peer properties like –peer computation time ($PC_T$), peer credibility ($C_P$), and distance between peers. The $C_P$ is probability that result produced by a peer is correct. $PC_T$ represent the time when a peer actually execute system's computations in presence of peer autonomy failure. In our proposed algorithm, we use above mentioned peer properties along with the turnaround time $D\ (T_D,\ T_P)$, to identify the group of most efficient peers. Suppose that total $M$ number of peers available in the system. The average peer credibility $C_V$ is given as:

$$C_{av} = (P_1.C_P + P_2.C_P + ..... + P_N.C_P) / M \tag{1}$$

The average peer computation $PC_V$ is given as:

$$PC_{av} = (P_1.PC_T + P_2.PC_T + ......+ P_N.PC_T) / M \tag{2}$$

The average turnaround time $D_V$ is given by:

$$D_{av} = \{P_1.D\ (T_D, T_P) + P_2.D\ (T_D, T_P) + .....+ P_N.D\ (T_D, T_P)\} / M \tag{3}$$

## 4.1 Protocol to Identify Most Efficient Peers

- A task distributor peer $T_D$, which has real time task for processing sends task availability message to all the peers connected to the system.

- A task processor peer $T_P$, upon receiving a task availability message, checks its availability, and in response sends a task request to $T_D$.

- Among all the peers responded only those peers will be selected in group of efficient peers if all the following conditions should be true for that peer:

    - The value of Peer Credibility $C_P$ must be one or more than required threshold value.

    - The value of peer computation time $PC_T$ must be greater than or equal to the average peer computation $PC_{av}$.

    - The value of turnaround time $D\ (T_D,\ T_P)$ must be laser than or equal to the average turnaround time $D_{av}$.

Those peers in the P2P computing system which are not satisfying the above conditions are not allowed to process the real time applications task.

**Algorithm of Peer Group Identification**

// $G_1$ : Group 1, $G_2$ : Group 2, $G_3$ : Group 3, $G_4$ : Group 4

// $PG_1$ : Peer Group 1, $PG_2$ : Peer Group 2, $PG_3$ : Peer Group 3, $PG_4$ : Peer Group 4

// $T_{P.}C_V$ : $C_V$ of $T_P$

// → : become a member of.

1. Start

2. If ($T_P$ ε $G_1$) then

3. If ($T_{P.}$ $C_V \geq \mu$) then

4. $T_P$ → $PG_1$;

5. Else

6. $T_P$ → $PG_3$;

7. Else if ($T_P$ ε $G_2$) then

8. If ($T_{P.}$ $C_V \geq \mu$) then

9. $T_P$ → $PG_2$;

10. Else

11. $T_P$ → $PG_4$;

12. Else if ($T_P$ ε $G_3$) then

13. $T_P$ → $PG_3$;

14. Else

15. TP → PG4;

16. End

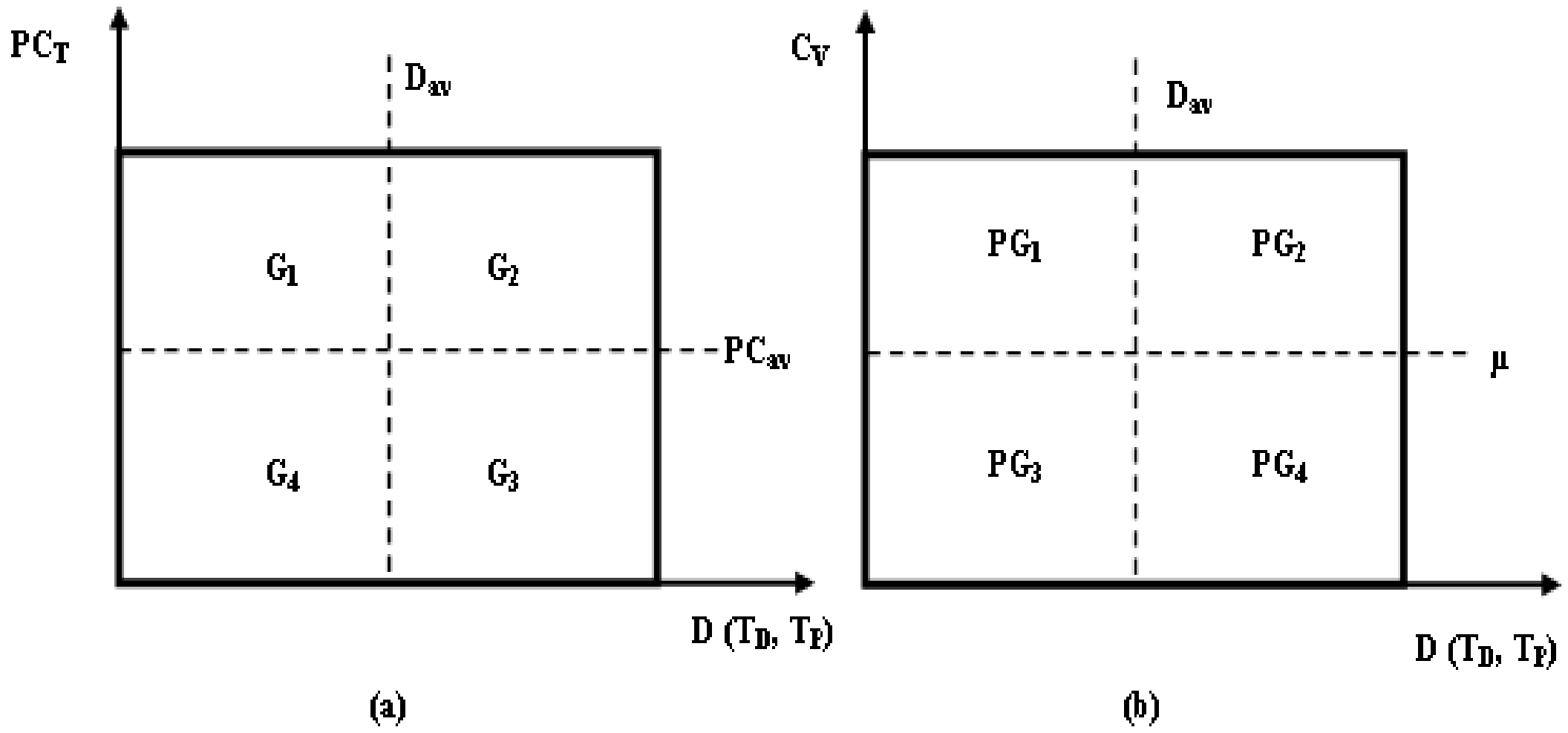

Figure 3. Peer Group Construction

The task processor peers are classified in different peer groups by the algorithm of peer group identification as given above. First, the responded task processor peers are classified into $G_1$, $G_2$, $G_3$, and $G_4$ classes depending on peer computation time $PC_T$ and peer distance $D\ (T_D, T_P)$ as shown in figure 3. Then, the classified task processor peers are classified into each peer group according to the peer credibility $C_V$. By the algorithm, peer groups are classified into four categories $PG_1$, $PG_2$, $PG_3$, and $PG_4$ as shown in fig. here $D_{av}$ is the average of distances and $\mu$ is the desired credibility threshold.

- The peer group '$PG_1$' represent a peer group in which all the peers have low values of $D(T_D, T_P)$ and high values of $PC_T$ and $C_V$. In group $PG_1$ all the peers have high possibility to successfully complete the task within the given deadline reliably because all the peers have $C_P$ greater than the required threshold value of $C_P$.

- The peer group '$PG_2$' has high values of $PC_T$, $C_V$, and $D(T_D, T_P)$. It has a high probability to produce correct results. However, it cannot complete the tasks within given deadline because distance between peers is high.

- The peer group '$PG_3$' has high probability to complete the task within given deadline because it has low values for $D\ (T_D, T_P)$, but it has less probability to produce correct results because the less values of $C_V$ and $PC_T$.

- The peer group '$PG_4$' has high $D\ (T_D, T_P)$ and less value for $C_V$ and $PC_T$. It has low probability to complete task within given deadline as well as produce correct results.

## 5. CONCLUSIONS

In this paper first we proposed a method to calculate the turnaround time of a task unit in P2P computing system at application level. We also give the algorithm for the proposed method. The values of turnaround time gives the overall impact of network conditions because these values are highly dependent on parameters like network bandwidth, task processor load and task processor peer's processing capability. The proposed method causes zero network traffic overhead, since it uses normal application traffic to measure values of turnaround time. For each measurement the task distributor require three indexed accesses to the lookup tables. However the proposed method does cause a slight memory overhead on task distributor peer. The task distributor peer must require storing a touple (task processor id, turnaround time) for every collaborated task processor peer.

The proposed method is capable of identifying a reliable group of most efficient peers on which execution of real time application task take placed. The peers are selected on the joint evaluation of peer computation time ($PC_T$), peer credibility ($C_P$), and turnaround time $D\ (T_D, T_P)$. The algorithm classifies peers in to four peer groups. The peer group '$PG_1$' is best suited to carry out the real time application tasks because peers in this group has the more credibility than the desired threshold value and also the distance $D\ (T_D, T_P)$ is also less.

## Authors

**Jigyasu Dubey**
Associate professor
Department of Information Technology
Shri Vaishnav Institute of Technology and Science Indore, (M. P.), India
Ph. D (Pursuing) (Computer Engineering) from DAVV, Indore
M.E. (Computer Engineering) in 2007 from DAVV, Indore
B.E. (Computer Science and Engineering) in 2000 from Vikaram University, Ujjain
***Areas of Interest:***
Computer Networks, Software Engineering, Object Oriented Analysis and Design

**Dr. (Mrs.) Vrinda Tokekar**
Professor & Head,
Department of Information & Technology,
Institute of Engineering & Technology,
Devi Ahilya University, Indore (M.P.) India
Ph. D. (Computer Engg.) in 2007 from DAVV, Indore
M.E. (Computer Engg.) in 1992 from DAVV, Indore
B.E. (Hons.) EEE, BITS Pilani in 1984,
***Areas of Interest:***
Computer Networks, Distributed Computing,, Security in Wireless Networks, e-Governance,
Multimedia Communication, Software Engineering